\documentclass[runningheads,fleqn]{svmult}

\usepackage{makeidx}   % allows index generation
\usepackage{graphicx}  % standard LaTeX graphics tool
\usepackage{subeqnar}  % subnumbers individual equations
\usepackage{multicol}  % used for the two-column index
\usepackage{physproc}  % flushleft layout of diverse elements,etc.
\makeindex             % used for the subject index
\begin{document}
\title*{Uniaxially anisotropic antiferromagnets in a field along the
  easy axis}
\toctitle{Uniaxially anisotropic antiferromagnets in a field}
% allows explicit linebreak for the table of content
%
%
\titlerunning{Uniaxially anisotropic antiferromagnets}
% allows abbreviation of title, if the full title is too long
% to fit in the running head
%
\author{W. Selke\inst{1}
  \and M. Holtschneider\inst{1}
  \and R. Leidl\inst{2}
  \and S. Wessel\inst{3}
  \and G. Bannasch\inst{1}}
\authorrunning{W. Selke et al.}
% if there are more than two authors,
% please abbreviate author list for running head
% 
% 
\institute{Institut f{\"u}r Theoretische Physik, RWTH Aachen, 52056
  Aachen, Germany
\and Department of Physics, Simon Fraser University, Burnaby, British
Columbia V5A 1S6, Canada
\and Institut f{\"u}r Theoretische Physik III, Universit{\"a}t
Stuttgart, 70550 Stuttgart, Germany}   

\maketitle              % typesets the title of the contribution
\begin{abstract}
Uniaxially anisotropic antiferromagnets in a field along the easy axis
are studied with
the help of ground state considerations and Monte Carlo
simulations. For classical models, the XXZ model as well as variants,
we analyze the role of non--collinear spin configurations of biconical
or bidirectional
type interpolating between the well--known antiferromagnetic
and spin--flop structures. Possible experimental applications
to layered cuprate antiferromagnets are discussed. Finally, results
of quantum Monte Carlo simulations for the $S$=1/2 XXZ model on a
square lattice are presented, and compared
with previous findings.
\end{abstract}
\section{Introduction}

\begin{figure}[b]
  \includegraphics[width=.7\textwidth]{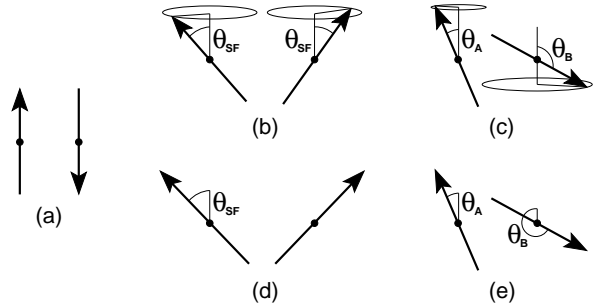}
  \caption{Spin orientations on neighboring sites showing AF (a), SF
    (b,d), and BC (c) as well as BD (e) ground state structures in XXZ
    (a,b,c) and anisotropic XY (a,d,e) antiferromagnets.
    }
  \label{eps1}
\end{figure}

Since many years uniaxially anisotropic antiferromagnets in a
field have been studied extensively, both experimentally and
theoretically. The magnets are known to display at low
temperatures, $T$, upon increasing the field $H$ along the
easy axis, antiferromagnetic (AF) and
spin--flop (SF) phases \cite{Neel}. A prototypical model describing
these phases is the XXZ Heisenberg antiferromagnet, with
the Hamiltonian

\begin{equation}
  {\cal H}_{\mathrm{XXZ}} = J \sum\limits_{i,j} 
  \left[ \, \Delta (S_i^x S_j^x + S_i^y S_j^y) + S_i^z S_j^z \, \right] 
  \; - \; H \sum\limits_{i} S_i^z
\end{equation}

\noindent
where $S_i^x$, $S_i^y$, and~$S_i^z$ denote
the spin components at lattice site~$i$. The
first sum runs over pairs of neighboring sites $(i,j)$ of
the square or cubic lattice; $J >0$ is the
exchange integral, and $\Delta$, $0 < \Delta < 1$, determines the
strength of the anisotropy along the easy axis ($z$--axis). The
field $H$ acts along the
$z$--axis. Classical XXZ antiferromagnets on square and cubic lattices have
been analyzed using Monte Carlo techniques first about three
decades ago \cite{BinLan,LanBin}.

Additional phases in the ($H,T$)--plane, observed in
experiments and theoretical studies, have been attributed
to, for instance, further anisotropy terms and/or
interactions ranging beyond nearest neighbors \cite{TM,LF,Wegner}.

One of the main aims of the present contribution is to draw attention to recent
theoretical analyses of the classical
XXZ antiferromagnet and its analogue for spins with
only two components, the anisotropic XY
antiferromagnet \cite{Holt1,Zhou1,Vic,Holt2,Holt3}. Especially, the
importance of non--collinear structures of biconical (BC) or
bidirectional (BD) type, see Fig. 1, is
emphasized \cite{Holt2,Holt3,Holt4}. By adding
a single--ion anisotropy term to the XXZ model, these structures may be 
enhanced or suppressed, depending on whether that term
favors a planar or a uniaxial anisotropy \cite{Holt3}. Some of the
recent theoretical analyses have been partly motivated by experiments on
quasi--twodimensional cuprate antiferromagnets, the 
'telephone number compounds' $(Ca,La)_{14}Cu_{24}O_{41}$
\cite{Ammer,Pokro,Mat,Rev2,Leidl2}. Thence, we
shall discuss also more complicated models for
uniaxially anisotropic twodimensional antiferromagnets proposed to describe
such compounds, in particular $Ca_5La_9Cu_{24}O_{41}$ (here, one may mention
previous and recent experimental studies 
on related quasi--twodimensional
antiferromagnets \cite{Bevaart,Gaulin,Cowley,Chris,Pini} as well).

\begin{figure}
  \includegraphics[width=.6\textwidth]{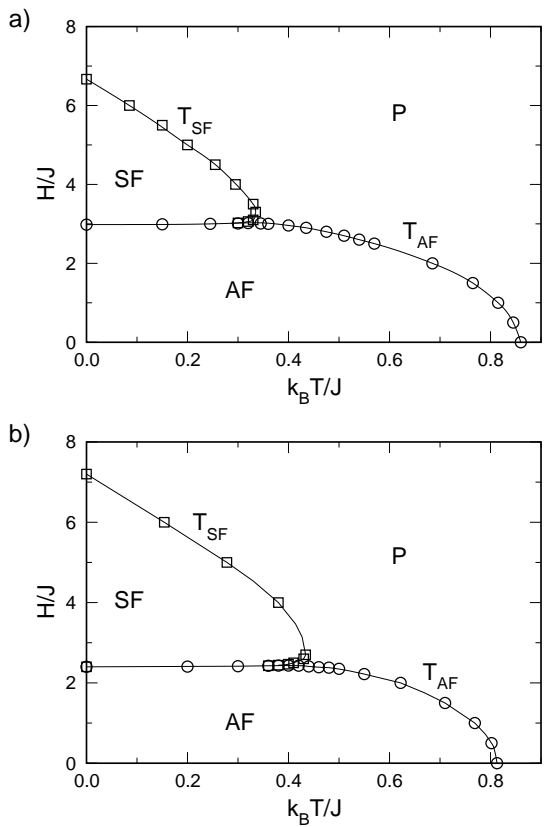}
  \caption{Phase diagram of the classical XXZ model with $\Delta$= 2/3
  (a) and 4/5 (b).}
  \label{eps2}
\end{figure}

Finally, we shall consider the quantum, $S=1/2$, version
of the XXZ antiferromagnet, which is equivalent to a 
Bose--Hubbard model \cite{Matsu,Schmid}, on a square
lattice, being, especially, of current interest in the context of
supersolids. The model is simulated using 
the method of stochastic series expansions (SSE). New results
\cite{Holt2} on the phase diagram will be compared
with previous ones \cite{Schmid,Kohno,Yuno}.

\begin{figure}
  \includegraphics[width=.6\textwidth]{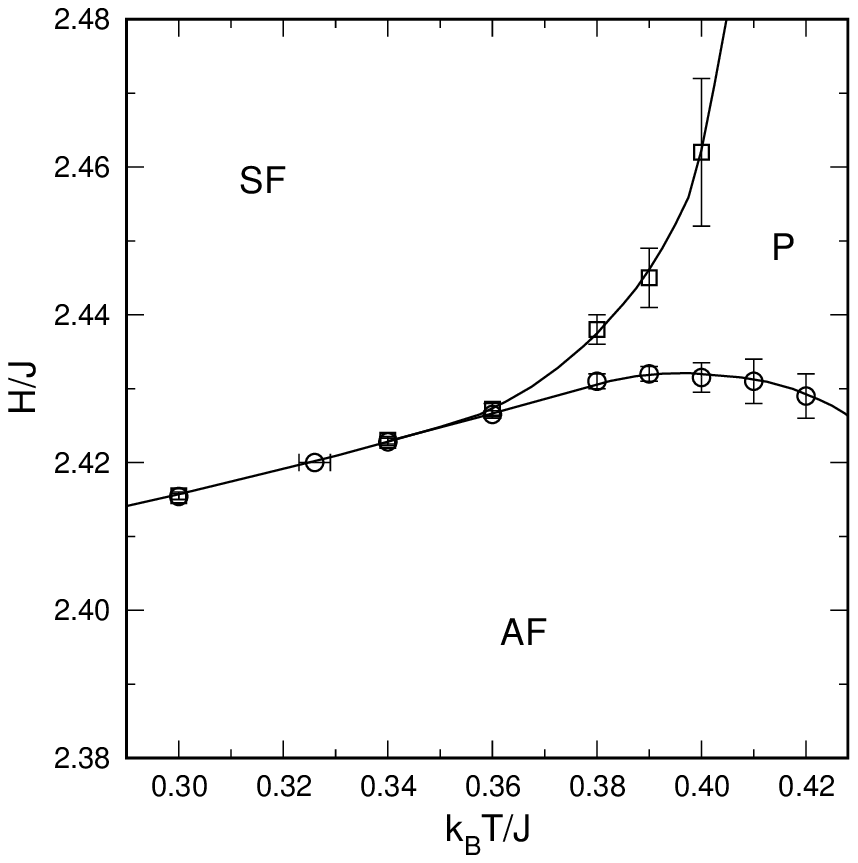}
  \caption{Phase diagram of the classical XXZ model near the maximum
    of the boundary line of the AF phase, $\Delta= 4/5$.}
  \label{eps3}
\end{figure}

Our findings will be summarized at the end of this contribution. 

\section{The classical XXZ Heisenberg antiferromagnet}

The classical XXZ Heisenberg antiferromagnet is described
by the Hamiltonian given in Eq. (1), where we shall deal with
spin vectors of length unity on square and
cubic lattices.

We first consider the twodimensional version. Phase
diagrams in the $(T,H)$--plane are depicted in Fig. 2, where we
set the exchange anisotropy $\Delta$ to be equal to 2/3 and 4/5, the
latter case being the standard
choice \cite{BinLan,LanBin,Holt1,Zhou1,Holt2}.

The general topology of the phase diagram seems
to be independent of the
concrete value of $\Delta$, $0 < \Delta <1$, comprising
the long--range ordered AF and the algebraically ordered
SF phase. The boundary lines to the disordered phase
are in the Ising universality class for the AF case, and in
the Kosterlitz--Thouless universality class \cite{KT} 
for the SF case \cite{BinLan,Holt1}. The
AF and SF phase boundary lines approach
each other very closely near the maximum of the SF phase boundary in the
$(T,H)$--plane, see Fig. 3. Accordingly, at low temperatures, there
may be either a direct transition between the AF and
SF phases, or two separate transitions with an extremely
narrow intervening phase may occur.

Indeed, recent simulations suggest a 
narrow (disordered) phase between the AF and SF
phases \cite{Holt1}, extending presumably
down to zero temperature \cite{Zhou1}. The evidence has been 
provided by determining the universality classes of the
transitions at low temperatures \cite{Holt1,Zhou1}, remaining to
be of Ising-- or Kosterlitz--Thouless type, and by
finite--size arguments in the limit of $T$ approaching
zero temperature \cite{Zhou1}. The presence of
that intervening phase may be argued \cite{Holt2} to be closely related to
the highly degenerate ground state occurring at the field
$H_{c1} = 4J \sqrt{1- \Delta^2}$ separating the AF
and SF structures. At that point, not only
AF and SF configurations have the same energy, but also
biconical structures. Those structures may be described
by the tilt angles $\Theta_A$
and $\Theta_B$ characterizing the orientations of
spin vectors at neighboring sites, belonging to the
two different sublattices $A$ and $B$, see Fig. 1. The two
tilt angles are interrelated by \cite{Holt2,Holt3}

\begin{equation}
 \Theta_B = \arccos \left( \frac{ \sqrt{1-\Delta^2} \; - \; \cos\Theta_A }{ 1 \; - \; \sqrt{1-\Delta^2} \cos\Theta_A } \right)
\end{equation}

\noindent
with the BC configurations interpolating continuously
between the AF and SF structures, where
the tilt angle $\Theta_A$ ranges from 0 to $\pi$.

The relevance 
of BC fluctuations in the transition region between the 
AF and SF phases at low temperatures may be conveniently
seen by studying probability
functions of the tilt angles, such as the
probability $p_2(\Theta_A,\Theta_B)$ for finding the
two angles, $\Theta_A$ and $\Theta_B$, at
neighboring sites and the probability $p(\Theta)$ for encountering the tilt
angle $\Theta$ \cite{Holt2,Holt3,Holt4}. An illustration
is depicted in Fig. 4, showing
that the line of local maxima in $p_2$ follows closely 
Eq. (2), signaling that the degenerate BC 
structures are present in that region. The rather low
probability for configurations deviating only slightly from 
the AF structure, see Fig. 4, is caused by the small radii of
their 'cones'. In
fact, at $T=0$, the probability of $p_2$ along the line given by Eq.(3) is
proportional to
$ \sqrt{sin(\Theta_A) sin(\Theta_B)}$. In any event, the, presumably, narrow
disordered phase seems to 
be governed by degenerate BC fluctuations, with a
hidden 'tetracritical point' at zero temperature \cite{Holt2}.

\begin{figure}
  \includegraphics[width=.6\textwidth]{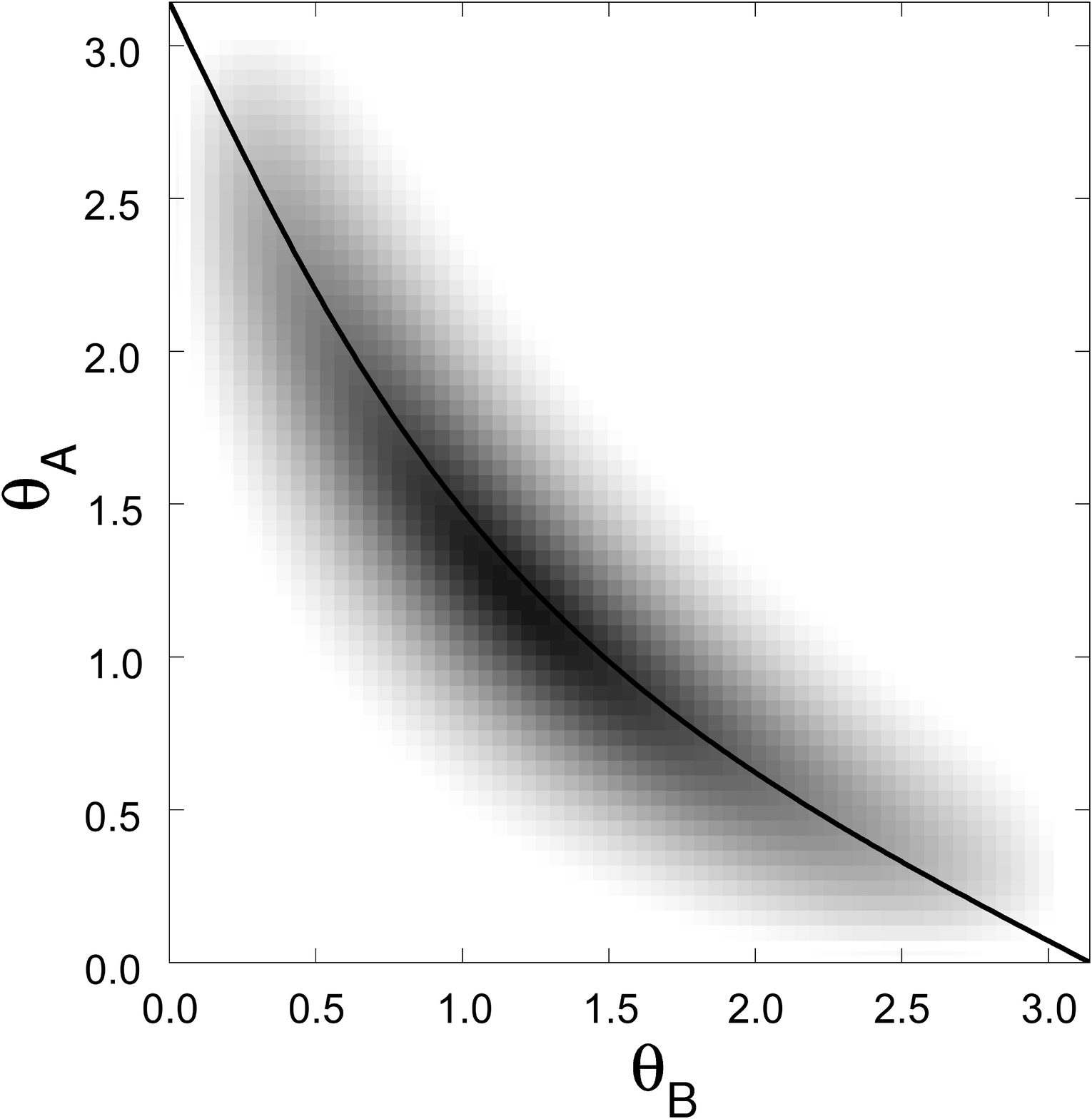}
  \caption{Probability ~\mbox{$p_2(\Theta_A,\Theta_B)$}
    showing the correlations between the tilt angles~$\Theta_A$
    and~$\Theta_B$ on neighboring sites for the
    XXZ antiferromagnet, with~$80 \times 80$
    spins, at~$H/J=2.41$, $k_BT/J=0.255$,
    and~$\Delta=\frac{4}{5}$. $p_2$~is proportional to
    the gray scale \cite{Holt2}. The
    superimposed black line depicts the relation between the two
    angles $\Theta_A$ and $\Theta_B$ in the ground state, Eq. (2).}
  \label{eps4}
\end{figure}

For the classical XXZ antiferromagnet on a cubic lattice early 
renormalization group arguments \cite{FN,KNF} and
Monte Carlo simulations \cite{LanBin} suggested that the 
triple point, at which the AF, SF and paramagnetic phases meet, is
a bicritical point with $O(3)$~symmetry (obviously,
such a point is excluded to occur, at $T>0$, in two dimensions due to the
well--known theorem by Mermin and Wagner \cite{MW}). Only a few years ago, 
this scenario has been questioned, based on high-order perturbative 
renormalization group calculations \cite{CPV}. It has been predicted
that, instead of the bicritical point, there may be a 'tetracritical 
biconical' \cite{KNF} point, due to an intervening ordered
'biconical' phase in between the AF and SF phases, or a point at which
first--order transition lines meet.

Our previous Monte Carlo simulations \cite{Holt1} for the
threedimensional XXZ antiferromagnet agreed with a first--order 
transition between the AF and SF phases at low
temperatures. Based on our current simulations \cite{BS}, again for 
$\Delta= 0.8$, we locate a triple point at $k_BT_t/J= 1.025 \pm
0.015$ and $H_t/J= 3.88 \pm 0.03$, in reasonable agreement
with the estimate by Landau and Binder \cite{LanBin}. So 
far we did not specify its (multicritical) character. Moreover, we analyzed 
$p_2$, showing that BC fluctuations occur
in the transition region between the AF and SF 
phases temperatures well below $T_t$. However, in contrast
to the twodimensional
case, one now observes a tendency towards coexistence of
the AF and SF phases, as reflected by fairly pronounced maxima at
the corresponding points in the $(\Theta_A, \Theta_B)$ plane, compare
with Fig. 4. We also estimated critical exponents from monitoring
the size dependence of the maxima in the longitudinal, i.e. along the easy 
axis, as well as the
transverse staggered susceptibilities and the specific heat near
the transition between the AF and SF phases at
temperatures somewhat below that triple point. In agreement
with previous
findings \cite{LanBin,Holt1} and the behavior of $p_2$, a transition
of first order between the AF and SF phases
is strongly suggested \cite{BS}.

\section{Variants and applications to layered cuprate magnets}

\begin{figure}
  \includegraphics[width=.6\textwidth]{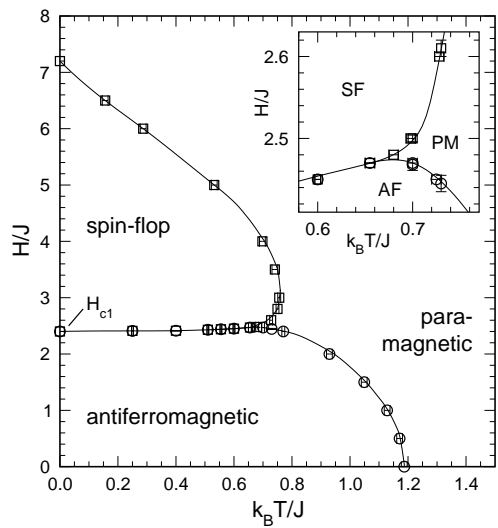}
  \caption{Phase diagram of the anisotropic XY antiferromagnet
   with $\Delta= 0.8$.}
  \label{eps5}
\end{figure}

In the following, classical variants of the twodimensional XXZ
antiferromagnet are studied. From a theoretical point of view, one may
like to check the robustness of the topology of the
phase diagram against
modifying the model, to identify genuine features. From
an experimental point of view, one may like to have a
(semi-)quantitative description of measurements. Addressing
the first aspect, we added a single--ion anisotropy term to the
XXZ model, and we also analyzed the anisotropic XY antiferromagnet
on a square lattice. Addressing
the experimental aspect, we focused
on layered cuprate antiferromagnets, the socalled
'telephone number compounds' $(Ca,La)_{14}Cu_{24}O_{41}$, in
particular $Ca_5La_9Cu_{24}O_{41}$.

\subsection{Classical anisotropic XY antiferromagnet}

Reducing the number of spin components to two and keeping the
uniaxial anisotropy, one arrives at the anisotropic XY
antiferromagnet, with the Hamiltonian

\begin{equation}
{\cal H}_{\mathrm{XY}} = J \sum\limits_{i,j} \left[ \, S_i^x S_j^x + \Delta
S_i^y S_j^y \, \right] \; - \; H \sum\limits_{i} S_i^x
\end{equation}
 
\noindent
where the $x$--axis is now the easy axis. As
before, $0 < \Delta < 1$. We set $\Delta= 0.8$. 

\begin{figure}
  \includegraphics[width=.6\textwidth]{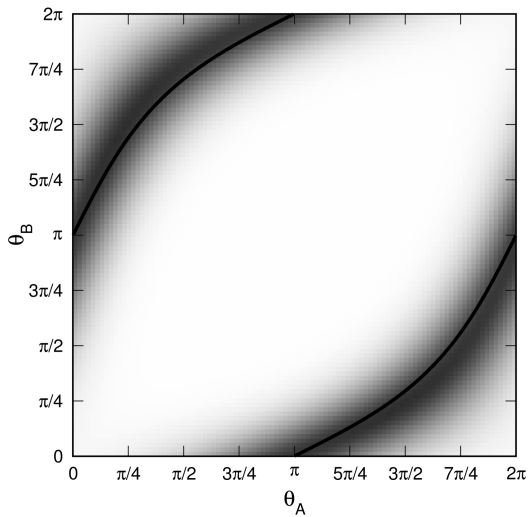}
  \caption{Probability $p_2(\Theta_A,\Theta_B)$ for
    the anisotropic XY antiferromagnet with $\Delta$= 0.8 for a system
    with $100 \times 100$ lattice sites in the transition region between
    the AF and SF phases at $k_BT/J= 0.558$ and
    $H/J=2.44$. $p_2(\Theta_A,\Theta_B)$ is proportional to the
    grayscale. The superimposed solid line depicts the relation between
    the two tilt angles $\Theta_A$ and $\Theta_B$ in the ground
    state, see Eq. (2).}
  \label{eps6}
\end{figure}

The topology of the phase diagram looks like in the XXZ
case \cite{Holt3,Holt4}, compare Figs. 2 and 3 with Fig. 5. The AF and SF
phase boundary lines approach
each other very closely near the maximum of the AF phase boundary in the
$(T,H)$--plane. Accordingly, at low temperatures, there
seems to be, again, either a direct transition between the AF and
SF phases, or two separate transitions with an extremely
narrow intervening phase.

In principle, now a bicritical point, with O(2) symmetry, may occur
at non--zero temperature, being of Kosterlitz--Thouless type. Our
simulations, however, suggest that, like in the XXZ case, there is a 
narrow disordered phase intervening between the AF and SF
phases down to temperatures well below the point where the
AF and SF phases approach each other very
closely \cite{Holt3,Holt4}. In particular, critical exponents
of the staggered susceptibilities are found to be compatible with
the Ising universality class holding now for the transitions of
the AF as well as the SF phases to the disordered phase. The
narrow intervening phase
seems to be due to degenerate bidirectional
structures, see Fig. 1, arising from the highly degenerate
ground state at the field $H_{c1}$, which separates the AF and
SF structures. This behavior is
in complete analogy to the one for the XXZ antiferromagnet.

At the highly degenerate ground state, the two tilt
angles $\Theta_A$ and $\Theta_B$ of
the BD structures are interrelated analogously
to Eq. (2). In contrast to the XXZ antiferromagnet, however, the
probability $p_2(\Theta_A,\Theta_B)$ along the line of
the interrelated tilt angles is now constant at $T=0$. Both
properties, the degeneracy and the (almost) constant value of $p_2$, tend
to hold at low temperatures as well. This is 
displayed in Fig. 6, depicting $p_2$ in the transition region
between the AF and SF phases at low temperatures.  

\subsection{Adding a single--ion--anisotropy to the XXZ model}

\begin{figure}
  \includegraphics[width=.6\textwidth]{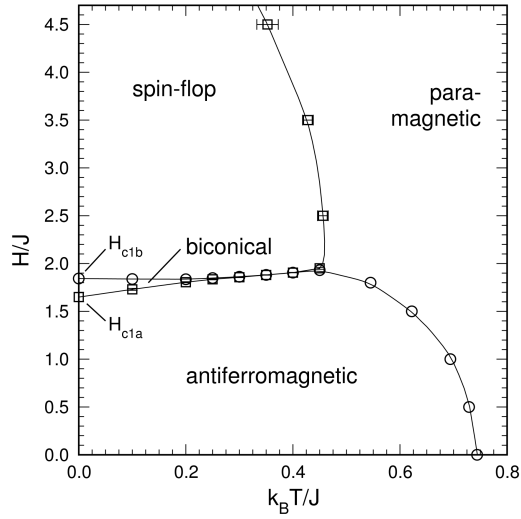}
  \caption{Phase diagram of the XXZ antiferromagnet with a competing 
   single--ion anisotropy, $\Delta= 0.8$ and $D/J= 0.2$.}
  \label{eps7}
\end{figure}

\begin{figure}
  \includegraphics[width=.6\textwidth]{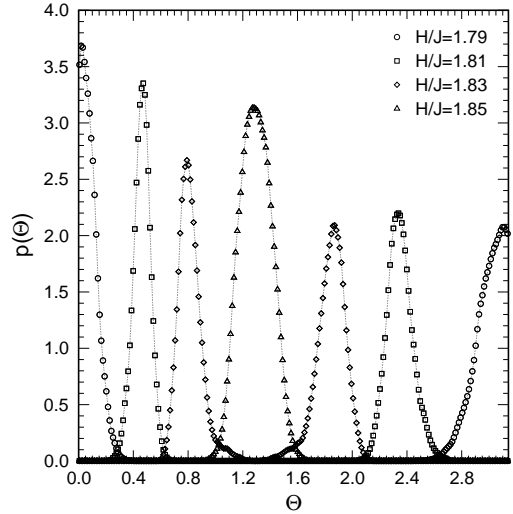}
  \caption{Histograms for the probability of the
tilt angle $p( \Theta)$ for the XXZ antiferromagnet with a competing
single--ion anisotropy, $D/J= 0.2$, at $k_BT/J= 0.2$, at the fields given
in the inset. Lattices with $80 \times 80$ spins are simulated.}
  \label{eps8}
\end{figure}

The classical XXZ model, Eq. (1), on a square lattice is modified by adding
a single--ion anisotropy term of the form

\begin{equation}
{\cal H}_{si} = D \sum\limits_{i} (S_i^z)^2
\end{equation}

\noindent
which either, $D < 0$, enhances the uniaxial
anisotropy $\Delta$, or, $D>0$, may weaken it due to a
competing planar anisotropy. The sign of $D$ will have 
drastic consequences for the phase diagram \cite{Holt3,Holt4,Holt5}. Due
to the single--ion term, the highly degenerate ground state, at 
$H_{c1}$ in the XXZ model, is removed, suppressing altogether the BC
structures, when $D < 0$, or 
spreading them over a finite range of fields, limited by
$H_{c1a}$ and $H_{c1b}$, when $D >0$, see Fig. 7.

In the latter case of a competing anisotropy, an
ordered BC phase arises at low temperatures, bordered by
the AF and SF phases, as shown in
Fig. 7. Based on renormalization group calculations
\cite{Bruce,Aharony,Mukamel,Domany}, the transition
between the BC and SF phases may be argued to be in the Ising universality
class, while the transition between the BC and AF phases
is expected to be in the XY universality class, being the
Kosterlitz--Thouless universality class in two
dimensions. This description is in accordance with our
simulational data, as inferred from critical exponents for staggered
susceptibilities and magnetizations at the two different phase boundary
lines \cite{Holt3,Holt4,Holt5}.

In the BC phase the interrelated tilt angles are
changing continuously, at fixed low temperature, with
the field. This behavior is displayed by
the probability function $p(\Theta)$, as illustrated
in Fig. 8. By increasing
the field, the peak positions correspond first to the AF
structure, shifting gradually towards each other, reflecting BC
structures, and finally merging
in one peak characterizing the SF phase.

As seen in Fig. 7, the extent of the BC phase shrinks with
increasing temperature. Eventually, the BC phase may terminate
at a tetracritical point \cite{LF,Bruce,Aharony,Mukamel,Domany}, where
the AF, SF, BC, and paramagnetic phases meet \cite{Holt3}.

\begin{figure}
  \includegraphics[width=.6\textwidth]{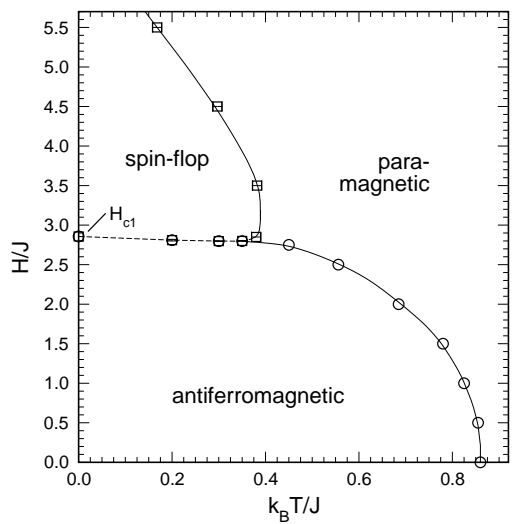}
  \caption{Phase diagram of the XXZ antiferromagnet with a single--ion
anisotropy fostering the uniaxiality, $\Delta= 0.8$ and $D/J= -0.2$.}
  \label{eps9}
\end{figure}

In the case of a negative single--ion
anisotropy, $D <0$, enhancing the exchange anisotropy, there are
no ground states of BC type. In Fig. 9, a
typical phase diagram is depicted, where $D/J= -0.2$, showing
long--range ordered AF and algebraically ordered
SF phases.

At low temperatures, we observe a transition of first order
separating the AF and SF phases. Evidence for that kind of
transition is provided, especially, by the critical exponent describing
the size--dependence of the maximum in the
longitudinal staggered susceptibility and by
a coexistence phenomenon of AF and SF structures in the transition
region between the two ordered phases, showing up, e.g., in
$p(\Theta)$ \cite{Holt3,Holt4}.

When the uniaxiality is solely due to a single--ion
anisotropy, $\Delta=1, D<0$, BC structures do not occur as
ground state. Accordingly, one may tend to believe that, at
low temperatures, a direct transition of first order between
the AF and SF phases takes place, in contrast to a conflicting
claim \cite{Costa}. Of course, this aspect needs to be
clarified.  

Here, attention is also drawn to interesting recent work
on a twodimensional Heisenberg antiferromagnet with long--range
dipolar interactions providing a uniaxial anisotropy \cite{Zhou2}.

\subsection{Descriptions related to $Ca_5La_9Cu_{24}O_{41}$}

The quasi--twodimensional uniaxially anisotropic Heisenberg antiferromagnet
$Ca_5La_9Cu_{24}O_{41}$ shows intriguing magnetic
features, as the consequence of an interplay of spin and charge
properties in the coupled $CuO_2$ spin
chains \cite{Ammer,Klingeler}, see below. Perhaps most
interestingly, at
low temperatures a sharp transition at a fairly low field along
the easy axis is followed, at a higher field, by an 
anomaly in the susceptibility.

\begin{figure}
  \includegraphics[width=.6\textwidth]{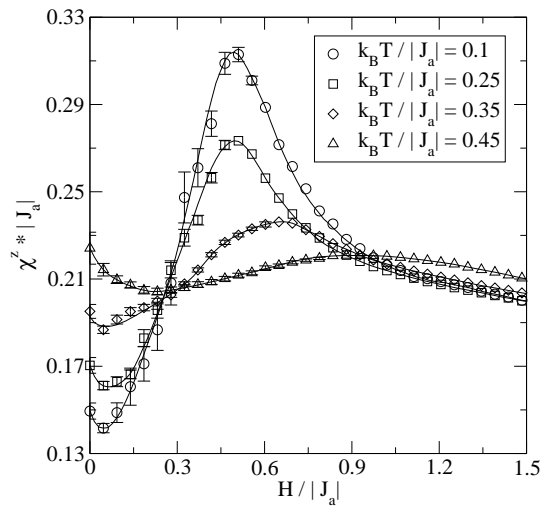}
  \caption{Susceptibility curves for different temperatures as
    simulated for a twodimensional uniaxially anisotropic Heisenberg
    antiferromagnet with quenched defects
    modeling $Ca_5La_9Cu_{24}O_{41}$ \cite{Leidl2}.}
  \label{eps10}
\end{figure}

To describe the measured spin--wave dispersion of
$Ca_5La_9Cu_{24}O_{41}$ Matsuda {\it et al.}
proposed a classical twodimensional model
with short--range exchange interactions and a
single--ion anisotropy \cite{Mat}. However, the model does
not reproduce thermal properties of that magnet like
the sharp transition \cite{Rev2,Leidl1a}. That transition has been
argued to indicate the thermal breaking of 'defect
stripes' \cite{Ammer,Klingeler,Kroll}. The 'defects' correspond
to nonmagnetic $Cu^{3+}$--ions, due to mobile holes, replacing some of the
magnetic $Cu^{2+}$--ions in the $CuO_2$ spin chains (the defect
concentration is about 10 percent). Indeed, such a
transition has been described by an
Ising model with mobile defects \cite{Pokro}. Introducing
nonmagnetic, mobile defects in the model of
Matsuda {\it et al.} is, however, not sufficient
to reconcile the discrepancy with
the measurements \cite{Leidl1}. 

The experimentally observed anomaly at higher fields has been
explained qualitatively as indicating the onset of merely $\it local$
spin--flop structures related to a significant decrease in the mobility
of the defects or holes \cite{Klingeler}. This presumption has
been used in a twodimensional uniaxially anisotropic 
Heisenberg antiferromagnet, with an exchange anisotropy and
short--range competing interactions by including
defects {\it quenched} at randomly chosen lattice sites \cite{Leidl2}. The
model parameters have been carefully chosen, partly on
theoretical, partly on experimental grounds. In fact, the model
then reproduces (semi--)quantitatively the field dependence of
the anomaly when changing the temperature, see Fig. 9. From
the simulations, one easily sees that the anomaly is driven by
the onset of merely local spin--flop structures, its local
character being due to the quenched random defects.

Note that the competing exchange interactions may induce helical
spin configurations when tuning the parameters
suitably \cite{Leidl2}.

We should like to mention a recent study on these 'telephone
number compounds' applying density functional theory \cite{Schwing},
which might also be useful to quantify model parameters.

%
%\subsubsection*{paragraph}
%

%This is a subsubsection, which we do not need
%

%
\section{$S=1/2$ XXZ quantum antiferromagnet on a square lattice}

Previous Monte Carlo simulations of the S=1/2 XXZ antiferromagnet
on a square lattice suggest that there is, at low temperatures, a
direct transition of first order between the AF and SF
phases \cite{Schmid,Kohno,Yuno}. 

\begin{figure}
  \includegraphics[width=.6\textwidth]{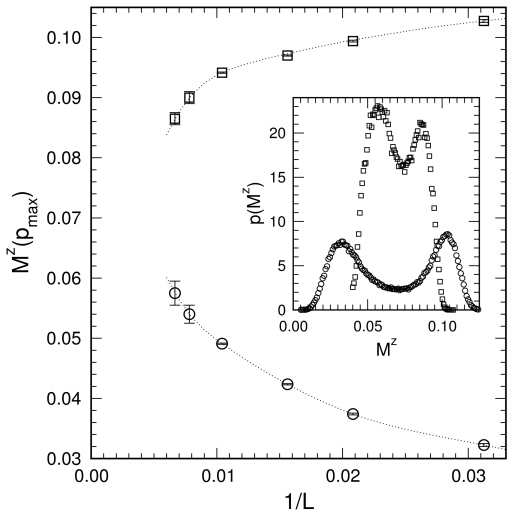}
  \caption{Positions of the maxima of the magnetization histograms as a
    function of the inverse system size, $1/L$, for the S=1/2 XXZ
    antiferromagnet on a square lattice, with $\Delta= 2/3$. The inset
    exemplifies two histograms
    for systems of linear size $L=32$~(circles) and $L=150$~(squares)
    at~$k_BT/J=0.13$ and the coexistence fields~$H/J=1.23075$
    and~$H/J=1.232245$ \cite{Holt2}.}
  \label{eps11}
\end{figure}

Schmid {\it et al.} \cite{Schmid} performed quantum Monte Carlo 
simulations to determine the phase diagram, for $\Delta=2/3$. They
found a topology which resembles that of the classical
XXZ antiferromagnet with a negative single--ion
anisotropy, compare with Fig. 9. On the boundary line of
the AF phase a tricritical point has been reported to
occur at $k_BT_{tc}/J \approx 0.141$, below which
the transition to the paramagnetic phase is of
first order. The triple point, at which the AF, SF, and disordered
phases meet, is proposed to be a critical endpoint, located
at $k_BT_{ce}/J \approx 0.118$. 

To check these predictions, we performed large scale quantum Monte
Carlo simulations \cite{Holt2} using the method of stochastic series
expansions with directed loop updates \cite{Sand}. From those
simulations, considering larger system
sizes and improved statistics, we conclude that the previous
analysis has to be viewed with care. For instance, we studied the
model, $\Delta= 2/3$, at $k_BT/J$= 0.13, i.e. in between
$T_{tc}$ and $T_{ce}$, near the AF phase
boundary. In particular, we monitored the size dependence
of peak positions in the
magnetization histograms, see Fig. 11. Obviously, the
two peaks, corresponding to AF and SF structures, may well 
coincide at the transition in the thermodynamic limit. Thence
the transition may
well be continuous, in contrast to the previous suggestion. Actually,
at the lowest temperature we studied, $k_BT/J \approx 0.096$, a continuous
transition might still occur \cite{Holt2}. We conclude that
the previous \cite{Schmid} scenario with the triple
point being a critical endpoint needs to be shifted to somewhat
lower temperatures, if it exits at all. 

In any event, a clue on possibly distinct phase diagrams for
the classical and quantum XXZ antiferromagnets on a square lattice
may be the possibly different role of biconical fluctuations. That
aspect deserves further studies.  

\section{Summary}

In this contribution we presented results of recent Monte Carlo
simulations on classical XXZ antiferromagnets in a field
along the easy axis as well as classical variants and on the
S=1/2 XXZ antiferromagnet on a square lattice.

Basic aspects of phase diagrams and applications to 
$Ca_5La_9Cu_{24}O_{41}$ are discussed.

The role of non--collinear structures of biconical and bidirectional
type in classical models is emphasized. These structures have
an important effect on phase diagrams, in particular, the transition
region between the AF and SF phases at low temperatures, and they
may provide a clue to explain the possibly different topology
of the phase diagrams of classical and quantum, S=1/2, XXZ
antiferromagnets in two dimensions.

We thank especially A. Aharony, K. Binder, B. B{\"u}chner,
B. Kastening, R. Klingeler, T. Kroll, D. P. Landau,
A. Pelissetto, V. L. Pokrovsky, and
E. Vicari for cooperation, help, remarks, information, and/or
encouragement. Financial suppport by the Deutsche
Forschungsgemeinschaft under grant SE324/4 and by JARA-SIM is
gratefully acknowledged. 

%
%\appendix
%
%\section*{Appendix}
%
%Appendices
%

%
\end{document}